\documentclass[12pt]{amsart}
\usepackage{amssymb, latexsym}
\def \C {{\mathcal C}}
\def \H {{\mathcal H}}
\def \O {{\mathcal O}}

\def \la {\lambda}
\def \al {\alpha}
\def \ga {\gamma}
\def \o  {\bf{ 0 }}
\def \ep {\epsilon}

\newcommand {\Fl} {{\mathbb F}}

\newtheorem{thm}{Theorem} 
\newtheorem{defn} {Definition} 
\newtheorem{lem} {Lemma} 
 
\newtheorem{cor} {Corollary} 
\newtheorem{rem} {Remark} 
 
\newtheorem{propn}{Proposition} 

\begin{document}

 \title[On Expander Graphs]{ On Expanders Graphs: Parameters and Applications}\thanks{  
{\large \bf Siam Journal of Discrete Mathematics}
 {\bf (submitted)}}
\author{H. L. Janwa \qquad and A. K. Lal}
\address{Department of 
Mathematics and Computer Science, University of Puerto Rico (UPR),
 Rio Piedras Campus,  P.O. Box: 23355,  San Juan, PR 00931 - 3355. }
\email{hjanwa@rrpac.upr.clu.edu}
\address{Department of Mathematics, Indian Institute of Technology Kanpur, Kanpur - 208016, INDIA}
\email{arlal@iitk.ac.in}
\thanks{The first author has his Research supported in part by NSF Grant No. CISE-9986985.}
\thanks{ The second author wishes to thank the University of Puerto Rico, Rio Piedras
Campus for the Visiting assignment. The work was completed during that period.}

\begin{abstract} 
We give a new lower bound on the   expansion coefficient of an edge-vertex   
graph of a $d$-regular graph.
As a consequence, we  obtain an  improvement on the lower bound on relative minimum distance
of the expander codes constructed by Sipser and Spielman.
We also derive  some improved results  on the vertex expansion 
of  graphs that help us in improving the
parameters of the    expander codes
of  Alon, Bruck, Naor, Naor, and Roth. 
\end{abstract} 
 \maketitle
\vspace{0.5in} 

\noindent {\bf Mailing address:} \\
Professor H. L. Janwa, \\ Department of Mathematics and Computer Science,\\
University of Puerto Rico, Rio Piedras Campus,  P O Box: 23355
San Juan, PR 00931 - 3355.  USA \\
\email{hjanwa@rrpac.upr.clu.edu}

\newpage 
 
\section{Introduction}
\label{sec:introduction}
The notions of expanders, magnifiers and concentrators were introduced
by Bassalygo and Pinsker~\cite{bassalygo:pinsker}, and Pippeneger~\cite{pippeneger}
 in  1970s in the study of
communication networks (in particular switching circuits).
The areas of applications of expander graphs in general 
have  grown vastly. To quote F.T. Leighton \cite{leighton:book}: ``Expander graphs
will become basic building blocks for the future high performance computing".
Expander graphs were  used
by Ajtai, Koml\'os and Szemer\'edi
in constructing an $\O(n\log n)$ sorting network
on sequential machines and a $c \log n$ sorting network
on parallel machines (see \cite{sarnak:book} for references).
They have huge number of applications in Computer Science, in Communication
Sciences, and in Mathematics;  we refer to
Leighton and Maggs~\cite{leighton:maggs} and
Janwa and Rangachari~\cite{janwa:rangachari} for details.

In coding theory applications of expander graphs go back to
the work of Tanner in the analysis and algebraic synthesis
of the low density parity check codes introduced by R.G. Gallager~\cite{gallager:thesis}.
   Recently  Sipser and Spielman~\cite{sipser:spielman}  constructed
 ``Expander  Codes" based on expander graphs. They also gave 
a fast decoding algorithm. Spielman~\cite{spielman}  further showed that
a class of expander codes can be encoded and  decoded in linear time.
Subsequent work by other researchers have shown that one can use these
ideas to construct codes that give performance near Shannon capacity!!
Thus they show performance better than the famous {\it turbo codes.}
Many recent researches address this question (see for example \cite{brian:rosenthal}).

Expander graphs  also form a basis of the work by Alon et al. \cite{alon:et}  in
construction of their expander codes which asymptotically attain the
Blokh-Zyablov bound.

The outline of the paper is as follows. 
Since the construction of Sipser and Spielman relies heavily on the edge-vertex graphs,
we derive  a new lower bound on the   expansion coefficient of the edge-vertex   
graph of a $d$-regular graph (in Section II, Theorem~\ref{thm:tanner:imp}). In Section III, we use this bound  
to give an  improvement on the lower bound on relative minimum distance
of the expander codes (Theorem~\ref{thm:sipser:spielman})  constructed by  Sipser and Spielman~\cite{sipser:spielman}. 
We show that this lower bound is in fact better, sometimes, by a factor of $3$ (see Remark~\ref{rem:imp}). 
In Section IV, we derive some results concerning the expander codes constructed by Alon et al.~\cite{alon:et}.
We first derive an improvement  ( Lemma~\ref{lem:alon:et:imp}) on the lower bound on $|\partial S|$ (defined below) of Alon et al.~\cite{alon:et}. 
This helps us in obtaining an improved bound (Corollary~\ref{cor:alon:raman})  on the minimum distance of the expander codes of Alon et al.

\subsection{Expander Graphs}
Let $G = (V, E)$ be a graph on $n$ vertices with adjacency matrix $A.$ The matrix $A$ is a real symmetric matrix and has $n$ real eigenvalues, say, $\la_0 \geq  \la_1 \geq  \cdots \geq  \la_{n-1}.$ Then define
$$\mu(G) = \max \{ \la_1, |\la_{n-1}| \}.$$

Throughout this paper $\mu$ will be used in place of $\mu (G)$ if the graph $G$ is understood from the context.

Let $G = (V, E)$ be a graph. Then for any vertex $v \in V,$ we define $\partial v = \{ u \in V : (u, v) \in E \},$  and for any subset $S$ of the vertex set, 
\begin{equation}
\partial S = \bigcup\limits_{v \in S} \partial v.
\end{equation}
 Note that for any subset $S $ of the vertex set $V, \; $ $\partial S $ can also be defined as $\partial S = \{ u \in V : S \cap \partial u \neq \emptyset \}.$

Let $\partial^*v = \partial v \setminus \{v\}$ and $\partial^* S = \bigcup\limits_{v \in S} \partial^* v \; (= \partial S \setminus S).$ 

\begin{rem}
Observe that for a bipartite graph $\partial^* S = \partial S.$ Note that if $d(v, S)$ denotes the distance of the vertex $v \in V$ from $S \subset V,$ then $$\partial^* S  = \{ v \in V : d(v, S) = 1 \},$$ i.e., $\partial^* S $ is the {\it boundary} of $S.$
\end{rem}

\begin{defn}[Expansion Coefficient]\label{defn:expander}
For $\al > 0,$ the {\it expansion coefficient,} denoted $c(\al),$ of a graph 
$G = (V, E)$ is defined as $$ c(\al) = \min \left\{ \frac{ |  \partial S | }{|  S |} : S \subset V, \; \;\;  0 < |  S| \leq \al | V| \right\} .$$
\end{defn}

\begin{defn}[$(n, c, d, \ep, \delta)$-Expanders]\label{defn:bipartite:expander}
A $(c, d)-$regular bipartite graph $G = (V, E)$ is called an 
$(n, c, d, \ep, \delta)-${\it{expander}}  if it is a graph on $n$ input vertices $I,$ in which every subset of at most $ \ep n$ vertices of $I$ expands by a factor of at least $\delta,$ {\it i.e.}, 
$$ \forall \;  S \subset I, \;{\mbox{ with }} | S|  \leq \ep n, \;\; | \partial^* S|  \geq \delta | S| .$$
\end{defn}

\begin{rem}
The above two definitions were motivated in the literature by the concept of a bounded strong concentrator  defined by Gabber and Galil~\cite{gabber:galil}.
\end{rem}

\begin{thm}[Tanner\cite{tanner}]\label{thm:tanner}
A $(c,d)-$regular graph $G$ on $n$ vertices is an expander graph with parameters $(n, c, d, \al, c(\al))$ and the expansion coefficient
\begin{equation}
c(\al) \geq \frac{c^2}{[ \al  c d  +  \mu^2 ( 1 - \al)  ]}.
\end{equation}
\end{thm}

\section{A Bound on the Expansion Coefficient of Edge-Vertex Graphs}

Let $G$ be a $d-$regular graph on $n$ vertices with $M$ as its edge-vertex incidence matrix. Then the {\it edge-vertex graph} of $G,$ denoted $\H,$ is a graph with adjacency matrix, $$A(\H) = \left[ \begin{array}{cc} \o & M \\ M^t & \o \end{array} \right].$$ The edge-vertex graph is a bipartite graph with $\displaystyle\frac{dn}{2}$ input vertices, each of degree $2,$ and $n$ output vertices, each of degree $d.$
If $A(G)$ is the adjacency matrix of the graph $G$ then using the fact  that the non-zero eigenvalues of $M M^t$ and $M^t M$ are same, we have the following lemma.

\begin{lem}
Let $G$ be a $d-$regular graph on $n$ vertices with $M$ as its edge-vertex incidence matrix. Then
$M^t M = A(G) + d I$ and $$\la_0(\H) = \sqrt{2 d} \;\; {\mbox{ and }} \;\; \mu(\H) = \sqrt{d + \mu(G)}.$$
\end{lem}

Hence, we use Theorem~\ref{thm:tanner} to conclude the following.

\begin{propn}\label{propn:dreg:bi}
Let $G$ be a $d-$regular graph on $n$ vertices. Then the edge-vertex graph $\H$ is an $(nd/2, 2, d, \al, c(\al))$ expander graph with $$\la_0(\H) = \sqrt{2 d} \;\; {\mbox{ and }} \;\; \mu(\H) = \sqrt{d + \mu(G)},$$ and 
\begin{equation} \label{eqn:bi:tanner}
c (\al) \geq \frac{4}{ \al(d - \mu) + (d + \mu)}.
\end{equation}
\end{propn}

We give an improvement on the bound in (\ref{eqn:bi:tanner}) using the following result of 
 Alon and Chung~\cite{alon:chung}.

\begin{lem}[Alon and Chung\cite{alon:chung}]\label{lem:alon:chung}
Let $G$ be a $d-$regular graph on $n$ vertices. Let $S $ be a subset of the vertex set of $G$ with $| S|  = \ga n.$ Then, for $e(S),$ the number of edges contained in the subgraph induced by $S$ in $G,$ 
\begin{equation}
\left|  e(S) - \frac{1}{2} \; d \; \ga^2 \; n \right|  \leq \frac{1}{2} \mu \ga (1 - \ga) n.
\end{equation}
\end{lem}
\begin{rem}
Observe that in Lemma~\ref{lem:alon:chung}, we can replace $\mu$ by $\la_1$ and $\la_{n-1}$ to get a better bound on $e(S),$ given by 
$$\frac{1}{2} d \ga^2 n  + \frac{1}{2} \la_{n-1} \ga (1 - \ga) n \leq e(S) \leq \frac{1}{2} \; d \; \ga^2 \; n + \frac{1}{2} \la_1 \ga (1 - \ga) n.$$
\end{rem}

\begin{rem}\label{rem:alon:chung}
Observe that Lemma~\ref{lem:alon:chung} means that if we have $$  e(S)   \geq \frac{1}{2} \; d \; \ga^2 \; n + \frac{1}{2} \mu \ga (1 - \ga) n = \frac{dn}{2} \left[\ga^2 + \frac{\mu}{d} \ga(1-\ga)\right] $$ then the corresponding subset $S$ of $ \; V$ will have $| S|  \geq \ga n,$ since  \linebreak[4] $ \displaystyle \frac{dn}{2} \left[\ga^2 + \frac{\mu}{d} \ga(1-\ga)\right] $ is a decreasing function of $\ga.$
\end{rem}

\begin{thm} \label{thm:tanner:imp}
Let $G = ( V(G), E(G) )$ be a $d-$regular graph on $n$ vertices. Let $\H$ be the corresponding
$(nd/2, 2, d, \al, c(\al))$ expander graph. Then
\begin{equation}\label{eqn:tanner:imp} 
c(\al) \geq \frac{4 }{ \mu + \sqrt{\mu^2 + 4 \al (d - \mu) d}}.
\end{equation}
\end{thm}
\begin{proof}
Let  $V(\H) = I(\H) \cup O(\H),$ where $I(\H)$ represents the set of input vertices and $O(\H)$ represents the set of output vertices of the graph $\H.$

For any subset $S$ of $V(G),$ an edge  $(u, v) \in e(S)$ if and only if $(u, v) \in E(G)$ with $u, v \in S.$ This is equivalent to saying that $(u, v) \in e(S)$ if and only if  there exists a vertex $\tau \in I(\H)$ such that  $(\tau, u), (\tau, v) \in E(\H)$ with $u, v \in S \subset O(\H).$ Therefore,
$$ |  e(S) |  = |  \{ \tau \in I(\H) : ( \tau, u), (\tau, v) \in E(\H),\; u, v \in S \subset O(\H) \} | .$$
Now let $T \subset I(\H)$ with $| T|  = \al |  I(\H) |  = \al \displaystyle\frac{dn}{2}.$ Then corresponding to this subset $T$ of $I(\H),$ we will get a set $S $ of $V(G)$ such that $ T  \subset  \; e(S) $
and $\partial T = S.$
 Suppose  $ \al = \ga^2 + \frac{\mu}{d} \ga(1-\ga) $  and
 $ T \subset I(\H)$ with   $ | T|  = \displaystyle\frac{dn}{2} \left[\ga^2 + \frac{\mu}{d} \ga(1-\ga) \right].$ Then  $ |  e(S) |  \geq \displaystyle\frac{dn}{2} \left[\ga^2  + \frac{\mu}{d} \ga(1-\ga)\right] $ and thus Remark~\ref{rem:alon:chung} implies that \linebreak[4]   
$| \partial T |  \geq \ga n = \displaystyle\frac{2}{\ga(d - \mu) + \mu} | T|. $ Hence, 
\begin{equation}\label{eqn:exp:e(S)}
c(\al) \geq \frac{2}{\ga(d - \mu) + \mu}. 
\end{equation}
The lower bound on $c(\al) $ can be expressed as a function of $\al$ using the following equalities:
\begin{eqnarray*}
& & \al =  \ga^2 + \frac{\mu}{d} \ga(1-\ga) \\
&\Longleftrightarrow & (d - \mu) \ga^2 + \mu \ga - \al d = 0 \\
&\Longleftrightarrow & (d - \mu) \ga + \mu = \frac{\mu}{2} + \frac{1}{2}\sqrt{ \mu^2 + 
4 \al d (d - \mu)}.
\end{eqnarray*}
The lower bound of $c(\al)$ in (\ref{eqn:tanner:imp}) now follows from (\ref{eqn:exp:e(S)}).
\end{proof}

\begin{rem}
We note that any $( nd/2, 2, d, \al, c(\al))$ expander graph, $H,$ can be interpreted as an edge-vertex graph of a graph $G$ in an obvious manner.
\end{rem}

\begin{cor}\label{cor:tanner:imp}
The bound on $c(\al)$ obtained for an $( nd/2, 2, d, \al, c(\al))$ expander graph in Theorem~\ref{thm:tanner:imp} is better than the one derived in Proposition~\ref{propn:dreg:bi}.
\end{cor}
\begin{proof}
It is enough to show that
 \begin{eqnarray*}
& & \frac{ 4 }{\al (d - \mu) + d + \mu}  < \frac{ 4 }{\mu + \sqrt{ \mu^2 + 4 \al d (d - \mu)}}\\
&\Longleftrightarrow  & \al (d - \mu) + d + \mu \; >  \;  \mu + \sqrt{ \mu^2 + 4 \al d (d - \mu)} \\
& \Longleftrightarrow &  [ d - \al(d  - \mu)]^2  \; > \; \mu^2 \\
& \Longleftrightarrow &  (d - \mu) ( 1 - \al) \; > \; 0
\end{eqnarray*}

which is always true since $1 > \al $ and $ d > \mu.$ 
\end{proof}

\begin{cor}\label{cor:expansion}
Let $G$ be a $d-$regular graph on $n$ vertices  and let \linebreak[4] $ d \ep > \mu.$ Then the corresponding edge-vertex expander graph $(nd/2, 2, d, \al_0, c(\al_0)),$ with $\al_0 = {{\ep (d \ep - \mu)} / {(d - \mu)}}$ has $$ c(\al_0) \geq \frac{2}{d \ep}.$$
\end{cor}
\begin{proof}
Using Theorem~\ref{thm:tanner:imp}, we know that 
\begin{equation} \label{eqn:expansion}
c(\al_0)  \geq \frac{4 }{ \mu + \sqrt{\mu^2 + 4 \al_0 (d - \mu) d}}.
\end{equation}
 Now replacing $ \al_0 (d - \mu)$  in the denominator by $\ep(d \ep - \mu)$ in (\ref{eqn:expansion}) we get the required result.
\end{proof}

Corollary~\ref{cor:expansion} is used in the next section to give an improved bound on the minimum distance of some expander codes of Sipser and Spielman~\cite{sipser:spielman}.

\section{Improved Bounds on the Parameters of \\Some Expander Codes}\label{sec:improvement:expcode}

\begin{defn}[Expander codes, Sipser and Spielman]\label{defn:expcodes}
Let $G$ be a \linebreak[4] $(c, d)-$regular bipartite graph between a set of $n$ input vertices $\{v_1, v_2, \ldots, v_n\},$ called {\it variables}, and a set of $\displaystyle\frac{n c}{d}$ output vertices $\{{\sf C}_1, {\sf C}_2, \ldots, {\sf C}_{nd/2} \},$ called {\it constraints}. Let $b(i,j)$ be a function designed so that, for each constraint ${\sf C}_i,$ the variables neighbouring ${\sf C}_i$ are $v_{b(i,1)}, \ldots, v_{b(i,d)}.$ Let $C$ be an error correcting code of block length $d.$ The {\it expander code} $\C(G,C)$ is the code of block length $n$ whose codewords are the words $(x_1, x_2, \ldots, x_n)$ such that for $ 1 \leq i \leq {nc/d},$ $(x_{b(i,1)}, \ldots, x_{b(i,d)})$  is a codeword of $C.$
\end{defn}

\begin{lem}[Sipser and Spielman~\cite{sipser:spielman}]\label{lem:sipser:spielman}
Let $G$ be an $(n, c, d, \al, c/{d \ep})$ expander graph and $C$ be an $[d, r d, \ep d]$ linear code. Then the expander code $\C(G,C)$ has rate at least $c r - (c - 1)$ and minimum relative distance at least $\al.$
\end{lem}

We  now  give an improvement to a result of Sipser and Spielman~\cite[Lemma 15]{sipser:spielman} 
on the minimum distance of expander codes. This is because we are able to derive a new
 lower bound on the expansion coefficient of an edge-vertex graph $\H$ of a $d-$regular graph 
$G$ (cf. Theorem~\ref{thm:tanner:imp}). This also helps in concluding that $d-$regular good 
expander graphs that are not bipartite also lead to good-expander codes (via the edge-vertex 
graphs).

\begin{thm}\label{thm:sipser:spielman}
Let $G$ be a $d-$regular graph on $n$ vertices and edge-vertex graph $\H$ with $d \ep > \mu.$ Also, let $C$ be an $[d, r d, \ep d]$ linear code. Then the expander code $\C(\H, C)$ has rate at least $2 r - 1$ and minimum relative distance at least $$\frac{\ep ( d \ep - \mu)}{d - \mu}.$$
\end{thm}
\begin{proof}
Corollary~\ref{cor:expansion} implies that the edge-vertex graph $\H$ is an \linebreak[4] $(\displaystyle\frac{nd}{2}, 2, d,\frac{\ep ( d \ep - \mu)}{d - \mu}, \frac{2}{d \ep})$ expander graph. Hence Lemma~\ref{lem:sipser:spielman} implies the required result.
\end{proof}

\begin{rem}\label{rem:imp}
Observe that our bound on the minimum relative distance of the expander code is better than the bound $ \left[ \displaystyle\frac{( d \ep - \mu)}{d - \mu} \right]^2$ 
obtained by Sipser and Spielman~\cite[Lemma 15]{sipser:spielman}. We achieve an improvement by a factor of  
\begin{equation}\label{eqn:imp}
1 + \frac{ \mu (1 - \ep)}{(d \ep - \mu)}.
\end{equation}\end{rem}
 As a concrete example, we consider the infinite sequence $\{ G_i \}_{i=1}^\infty$ of  \linebreak[4] $(q+1)-$regular 
Ramanujan graphs constructed by Morgenstern~\cite{morgenstern} (his work was  inspired by the 
work of Lubotzky, 
Phillips and Sarnak~\cite{LPS} and that of Margulis~\cite{margulis}). Let us choose $q = 2^{2 m}$ for some $m$ and 
$\epsilon \approx \displaystyle \frac{3 \cdot 2^m}{(2^m + 1)^2};$ then  
$\mu(G_i) \approx 2^{m+1}  $ for every 
$i = 1, 2, \ldots, \infty.$  For the initial code, we  choose a binary subfield subcode of the 
extended Reed-Solomon code of length $2^{2 m} + 1$ with the minimum distance at least $ \epsilon (2^{2 m} + 1).$ 
Then (\ref{eqn:imp}) implies  that we get an improvement by a factor of $3$ on the relative minimum distance
of the infinite sequence of codes $\{ \C(G_i, C) \}_{i=1}^\infty.$

\section{ Improved Bounds on the Minimum Distance of Expander Codes Introduced by Alon {\it et al.}}

We now give an improved  bound on the relative distance of the expander codes defined  by Alon {\it{et al.}}~\cite{alon:et}. 

\begin{lem}\label{lem:alon:et}
Let $G = (V , E)$ be a $d-$regular graph on $n$ vertices. Then, for all subsets $S$ of $ V $  with $| S|  = \al n,$ 
\begin{equation}
\sum_{v \in V} |  S \cap \partial v| ^2 \leq \left[\al (d^2 - \mu^2) + \mu^2 \right] | S| .
\end{equation}
\end{lem}
\begin{proof}
Let $\{ \phi_i \}_{i=1}^n$ be the orthonormal set of eigenvectors of the real symmetric adjacency matrix $A$ of the graph $G,$ with corresponding eigenvalues $\la_0 \geq \la_1 = \mu \geq \la_2 \geq \cdots \geq \la_{n-1}.$ Then by Perron-Frobenius theorem, the largest eigenvalue of $A$ is $\la_0 = d$ with the corresponding eigenvector $\phi_0 = (1, 1, \ldots, 1)^t/\sqrt{n}.$

Let $f = f_{S}$ be the characteristic vector of the set $S.$ 
Let $ f = \sum\limits_{i=0}^{n-1} \ga_i \phi_i.$ Then, $\ga_0 = \langle f, \phi_1 \rangle = \al \sqrt{n},$ and 
\begin{eqnarray*}
\sum_{v \in V} |  S \cap \partial v| ^2 & = & \langle A f, A f \rangle 
= \sum_{i=0}^{n-1} \la_i^2 \ga_i^2 \\
& \leq & \la_0^2 \ga_0^2 + \la_1^2 ( \ga_1^2 + \cdots +  \ga_{n-1}^2)  \\
&=& \ga_0^2 (\la_0^2 - \la_1^2 )  + \la_1^2 \langle f, f \rangle \\
&=& \al^2 n (d^2 - \mu^2) + \mu^2 \al n \\
&=&  [\al (d^2 - \mu^2) + \mu^2] | S| . 
\end{eqnarray*}

\end{proof}

 Lemma~\ref{lem:alon:et} helps us  prove the following lower bound on $\partial S, $ for $S \subset V.$

\begin{lem}\label{lem:alon:et:imp}
Let $G = (V , E)$ be a $d-$regular graph on $n$ vertices. Let $S \subset V$ with $| S|  = \al n. $ Then 
\begin{equation}
|  \partial S |  \geq \frac{d^2 | S| }{[\al (d^2 - \mu^2) + \mu^2]},
\end{equation}
and 
\begin{equation}\label{eqn:bar}
| \partial {\overline{  S}} |  = | \{ v \in V : S \cap \partial v = \emptyset \} |  \leq \frac{\mu^2 | \overline{S}| }{[\al (d^2 - \mu^2) + \mu^2]}.
\end{equation}
\end{lem}
\begin{proof}
It can be seen that $\sum\limits_{v \in V} |  S \cap \partial v|  = d | S|  = d \al n.$ Hence using Lemma~\ref{lem:alon:et} and  the Cauchy-Schwartz inequality, 
 we get  
$$ 
\frac{d^2 \al^2 n^2}{| \partial S| } = \frac{\left( \sum\limits_{v \in \partial S} | S \cap \partial v|  \right)^2}{ |  \partial S |  } \leq \sum_{v \in V} |  S \cap \partial v| ^2  
\leq [\al (d^2 - \mu^2) + \mu^2]  \al n.
$$
Thus, $$|  \partial S |   \geq \frac{d^2 \al n}{[\al (d^2 - \mu^2) + \mu^2]} = \frac{d^2 | S| }{[\al (d^2 - \mu^2) + \mu^2]}.$$ The bound in (\ref{eqn:bar}) on $|  {\partial \overline S} |$ follows by observing that  $|  \partial{\overline S} | = n - |  \partial S|.$ 
\end{proof}

Lemma~\ref{lem:alon:et:imp} gives a better bound on $\partial{ \overline S}$  as compared to the bound obtained by Alon {\it{et al.}}~\cite{alon:et} which is $\partial{ \overline S} \leq \displaystyle \frac{\mu^2 | \overline{S}| }{\al d^2 }.$ Therefore, we obtain a stronger bound on the minimum weight of expander codes of Alon {\it et al.} as compared to the bound obtained by 
them in~\cite{alon:et}. The definition of their expander codes follows.

\begin{defn}[Expander Codes, Alon {\it et al.}~\cite{alon:et}]\label{defn:expandercode}
Let $G$ be a $d-$regular expander graph on $n$ vertices. Let $C$ be an $[n, k, \delta_0 n ]$ code over $\Fl_q.$ Then Alon {\it{et al.}} use $G$ to define an {\it expander code} $\C_{\mbox{exp}}$ over $\Fl_{q^d}$ as follows: Label the vertices of $G$ by $1, 2, \ldots, n.$ For each vertex $i \; (1 \leq i \leq n),$ let $l_1(i), l_2(i), \ldots, l_d(i)$ denote the $d$ vertices that are adjacent to $i,$ indexed according to some prespecified order.

The expander map $\phi_{\mbox{exp}} : \Fl_q^n \longrightarrow \Fl_{q^d}^n$ is defined as follows: \\ $[a_1, a_2, \ldots, a_n] \longrightarrow [\al_1, \al_2, \ldots, \al_n],$ where $\al_i = [a_{l_1(i)}, a_{l_2(i)}, \ldots, a_{l_d(i)}]$ (by identifying $\Fl_{q^d}$ and $\Fl_q^d).$

The expander code $\C_{\mbox{exp}}$ is defined to be $\phi_{\mbox{exp}}(C).$ 
\end{defn}
\begin{rem}\label{rem:alon}
It was observed by Alon {\it et al.} that 
\begin{description}
\item i) $\;$ the expander code $\C_{\mbox{exp}}$ may no longer be a linear code.
\item ii) $ \;$  the Hamming distance between any two codewords $c_1, c_2 \in \C_{\mbox{exp}}$  equals the Hamming weight of $c_1 - c_2.$
\end{description}
\end{rem}

\begin{rem} We further observe that 
\begin{description}
\item i) $\;$ the expander code $\; \C_{\mbox{exp}}$ is $\Fl_q-$linear.
\item ii) $\;$ the $\Fl_q$ image of $\; \C_{\mbox{exp}}$  is a special permutation of the $d-$repetition of the original code $C.$
\end{description}
\end{rem}

\begin{thm}\label{thm:raman}
Let $G$ be a $d-$regular graph on $n$ vertices and  let $C$ be an $[n, r_0 n, \delta_0 n]$ code over $\Fl_q.$ Then $\C_{\mbox{exp}}$ is a code of length $n,$ rate $\bigl(\displaystyle\frac{r_0}{d} \bigr) n$  over $\Fl_{q^d}$ with minimum distance at least  $$\frac{d^2 \; \delta_0 \; n}{ \delta_0 (d^2 - \mu^2) + \mu^2}.$$
\end{thm}
\begin{proof}
Let $S = {\mbox{supp }} (u)$ for any codeword $u = [a_1, a_2, \ldots, a_n]  \in C$ with ${\mbox{wt}}(u) = \delta_0 n.$ Let $U \in \C_{\mbox{exp}}$  be the corresponding  codeword. Then $ i \in {\mbox{supp }}(U)$ if and only if there exists $j, \; 1 \leq j \leq d$ such that $ a_{\ell_j(i)} \neq 0.$ That is, if and only if $\; \partial \{i\} \cap S \neq \emptyset.$ Hence $\;\; {\mbox{supp }}(U) = \partial S.$ Therefore the result follows by using Lemma~\ref{lem:alon:et:imp} and Remark~\ref{rem:alon}.
\end{proof}

\begin{cor}\label{cor:alon:raman}
Let $G$ be a $\Delta-$regular Ramanujan graph on $n$ vertices and let $C$ be as in Theorem~\ref{thm:raman}. Then $\C_{\mbox{exp}}$  has minimum distance at least $$\frac{\Delta  \; \delta_0 \; n}{ \delta_0  \Delta + 4(1 - \delta_0)}.$$
\end{cor}
\begin{proof}
Using Theorem~\ref{thm:raman} we derive that the minimum distance of $\C_{\mbox{exp}}$  is at least $\displaystyle\frac{\Delta^2  \; \delta_0 \; n}{ \delta_0 ( \Delta^2 - \mu^2) + \mu^2}.$ But for Ramanujan graphs $\mu^2 \leq 4 (\Delta -1) \leq 4 \Delta.$ Hence the result follows by replacing $\mu^2 $ by $4 \Delta.$
\end{proof}

\begin{rem}
Our bound  on the minimum distance of 
$\C_{\mbox{exp}}$ is greater than \linebreak[4] $ \displaystyle \frac{\left[ \delta_0 \Delta^2 - \mu^2(1 - \delta_0) \right]}{\delta_0 \Delta^2} n, $ the one given by 
 Alon {\it et al.}~\cite[Lemma 1]{alon:et}.
\end{rem}

\vspace{0.25in}
{\it Acknowledgements:}  The second author would like to thank the Department of Mathematics and Computer Science, 
University of Puerto Rico, Rio Piedras Campus, San Juan, for their hospitality and  a  visiting appointment 
during August 2000 to July 2001, when the work was done.

\end{document}